\documentclass[superscriptaddress,%
 amsmath,amssymb,%
 aps,pra,twocolumn,%
floatfix,
]{revtex4-2}

\usepackage{graphicx}%
\usepackage{bm}%
\usepackage[colorlinks=true,%
bookmarks=false,%
linkcolor=blue,%
urlcolor=blue,%
citecolor=blue,%
breaklinks]{hyperref}

\usepackage{xcolor}
\usepackage{physics}

\begin{document}

\title{Spatial search by continuous-time quantum walks on renormalized Internet networks}

\author{Joonas Malmi}
\affiliation{QTF Centre of Excellence, Department of Physics, Faculty of Science, University of Helsinki, FI-00014 Helsinki, Finland}
\affiliation{InstituteQ - the Finnish Quantum Institute, University of Helsinki, FI-00014 Helsinki, Finland}
\affiliation{Algorithmiq Ltd, Kanavakatu 3 C, FI-00160 Helsinki, Finland}

\author{Matteo A. C. Rossi}
\affiliation{Algorithmiq Ltd, Kanavakatu 3 C, FI-00160 Helsinki, Finland}
\affiliation{QTF Centre of Excellence, Department of Applied Physics, Aalto University, FI-00076 Aalto, Finland}
\affiliation{InstituteQ - the Finnish Quantum Institute, Aalto University, FI-00076 Aalto, Finland}

\author{Guillermo Garc\'{i}a-P\'{e}rez}
\affiliation{QTF Centre of Excellence, Department of Physics, Faculty of Science, University of Helsinki, FI-00014 Helsinki, Finland}
\affiliation{InstituteQ - the Finnish Quantum Institute, University of Helsinki, FI-00014 Helsinki, Finland}
\affiliation{Algorithmiq Ltd, Kanavakatu 3 C, FI-00160 Helsinki, Finland}
\affiliation{Complex Systems Research Group, Department of Mathematics and Statistics,
University of Turku, FI-20014 Turun Yliopisto, Finland}

\author{Sabrina Maniscalco}
\affiliation{QTF Centre of Excellence, Department of Physics, Faculty of Science, University of Helsinki, FI-00014 Helsinki, Finland}
\affiliation{InstituteQ - the Finnish Quantum Institute, University of Helsinki, FI-00014 Helsinki, Finland}
\affiliation{Algorithmiq Ltd, Kanavakatu 3 C, FI-00160 Helsinki, Finland}
\affiliation{QTF Centre of Excellence, Department of Applied Physics, Aalto University, FI-00076 Aalto, Finland}
\affiliation{InstituteQ - the Finnish Quantum Institute, Aalto University, FI-00076 Aalto, Finland}

\date{\today}

\begin{abstract}
We study spatial search with continuous-time quantum walks on real-world complex networks. We use smaller replicas of the Internet network obtained with a recent geometric renormalization method introduced by García-Pérez et al., Nat. Phys. 14, 583 (2018). This allows us to infer for the first time the behavior of a quantum spatial search algorithm on a real-world complex network. By simulating numerically the dynamics and optimizing the coupling parameter, we study the optimality of the algorithm and its scaling with the size of the network, showing that on average it is considerably better than the classical scaling $\mathcal{O}(N)$, but it does not reach the ideal quadratic speedup $\mathcal{O}(\sqrt{N})$ that can be achieved, e.g. in complete graphs. 
However, the performance of the search algorithm strongly depends on the degree of the nodes and, in fact, the scaling is found to be very close to optimal when we consider the nodes below the $99$th percentile ordered according to the degree.
\end{abstract}

\maketitle

\section{Introduction}
\label{sec:introduction}
Continuous-time quantum walks (CTQWs), initially proposed in \cite{FarhiGutmann}, are the quantum analogues of continuous-time classical random walks, which describe the propagation of a particle over a discrete set of positions. Together with their discrete-time counterpart \cite{Aharonov1993}, they have received a lot of attention for their applications in quantum information processing \cite{Kempe2003,VenegasAndraca2012}, quantum computation \cite{Childs2009}, and quantum transport \cite{Muelken2011}. In the recent years, great progress has been made in the experimental implementation of CTQWs on different topologies, such as 2D lattices \cite{Poulios2014,Tang2018} and fractal graphs \cite{Xu2021}, and in their simulation on quantum devices \cite{Qiang2021,Qu2022deterministic}.

One of the most interesting algorithmic applications of CTQWs is spatial search \cite{ChildsGoldstone}, which is the problem of finding a marked element in a structured database. Essentially, this is the generalization, or the ``analog analogue'' \cite{analoganalogue}, of Grover's algorithm, where the spatial structure of a connected database has to be taken into account, and is usually encoded in a set of nodes and links between them, i.e., a graph.

Childs and Goldstone showed that a spatial search algorithm based on CTQWs \cite{ChildsGoldstone} can solve the spatial search problem in certain regular graphs with a quadratic speedup ($T = \mathcal{O}( \sqrt{N})$) when compared to the best possible classical algorithms ($T = \mathcal{O} ( N)$) (with $T$ being the total time and $N$ being the number of nodes in the graph).

Later research has shown that spatial search by CTQW is optimal for other graph topologies such as the star graph \cite{Novo2015,Cattaneo}, graphs with broken links \cite{Novo2015}, fractal graphs \cite{Agliari},  Erd\H{o}s-R\'enyi graphs \cite{SSoptimalforalmostallgraphs}, and 1D graphs with long-range interactions \cite{Lewis2021}. However, general conditions for a graph to be efficiently searchable are not known. It was shown that global symmetry and high connectivity are not necessary conditions for optimality of spatial search \cite{symmetryUnnecessary,connectivity}. Recently, an approximate prediction of the search time and success rate has been derived based on spectral properties of the graph \cite{optimality} (although as we will show later, the approximation may not hold for all graph topologies).

The networks found in real systems (e.g.~biological systems, communications systems, social  networks, etc.) are characterized by non-trivial topological features that depart from completely regular as well as purely random graphs and they are called complex networks. They have recently received large attention from the quantum physics community \cite{classicaltoquantum}, and in particular, they are studied in the context of CTQWs for coherent transport \cite{Muelken2011}, for example in biological systems \cite{Benedetti2019,Chisholm2021}, community detection \cite{faccin2014communitydetection}, and link prediction  \cite{moutinho2021quantum,goldsmith2022link}.

While spatial search has been extensively studied for regular and random graphs, quite little work has been done on studying its efficiency on complex networks, with very recent results obtained for synthetic networks, such as Kronecker graphs \cite{Kroneckergraphs} and Bollob\'as networks \cite{spatialsearchBollobas}, that reproduce some characteristics of complex networks, such as being small-world and scale-free. While synthetic models allow for fine control of the parameters of the network and to derive analytical results, they often fail at capturing the properties of real-world networks.

A crucial aspect when studying spatial search is the scaling of the average search time with the size of the network. 
While synthetic models allow for the generation of networks with an arbitrary number of nodes, with real networks one is limited to the observed ones.
In order to obtain replicas of the network with different sizes, in this paper we employ a recently proposed technique based on the geometric renormalization group \cite{geometricrenormalization,guillepnas}. The method takes the original complex network and uses geometric scaling to renormalize it while preserving the overall structure of the network. A real-space renormalization group \cite{Pathria}, originally used to study phase transitions in statistical mechanics, has been used to research complexity bounds of a spatial search algorithm in fractal networks, albeit with coin-based discrete-time quantum walks \cite{Boettcher}. However, this is different to the geometric renormalization group used in this paper, which is targeted towards complex networks.

In this paper, we study the effectiveness of spatial search by CTQWs on real networks by considering renormalized replicas of manageable size, namely, we consider the Internet network at the level of autonomous systems \cite{Internetnetwork, internetmapping}, and compare the dynamics on its renormalized replicas with the one on synthetic scale-free graphs (using the Barab\'asi-Albert model \cite{Barabasi-Albert}) and random graphs (using the Erd\"os-R\'enyi model \cite{Erdos-Renyi}) of similar sizes. Note that, while here we focus on the Internet as a paradigmatic example, the approach is general, as demonstrated in Ref. \cite{geometricrenormalization}. Remarkably, it has been shown that the same replicas can be used to study the inverse process, namely scaling up of real networks, through a process called geometric branching growth \cite{guillepnas}. Specifically, in our case, the optimality and scaling analysis on replicas allow us to infer the CTQW behavior on the original Internet network, and to predict its properties as the Internet grows.  

Summarizing, the geometric renormalization of complex networks, used here for the first time in the quantum context, has two advantages. On the one hand, it solves the issue of having just one instance of the network, as opposed to the case of synthetic networks. On the other hand, it overcomes the problem of the network size, which would be generally too large to be realistically studied. In this sense our results pave the way for the investigation of scaling and optimality of CTQW, and its algorithmic and transport applications, to any real network.

In the following, we show that, for the Internet network, the efficacy of the spatial search  depends on the degree of the target node, with large-degree nodes being characterised by the lowest success probability. Also,
the optimal value of the coupling constant, leading to the best scaling of the spatial search, depends on the degree of the target node, and specifically it is inversely proportional to it. When averaged over all target nodes, quantum spatial search on the Internet exhibits a better scaling than the classical $\mathcal{O}(N)$, but does not reach the optimal quadratic speedup $\mathcal{O}(\sqrt{N})$. However, when averaged over $99\%$ of the smallest-degree nodes, the scaling becomes very close to optimal.

This paper is structured as follows: in Sec.~\ref{sec:models} we review the relevant models used in this paper, namely the spatial search by CTQW and the geometric renormalization technique, as well as the numerical methods used in the simulations. In Sec.~\ref{sec:results} we present our results on the behavior of the spatial search algorithm on different layers of the renormalized Internet network and the corresponding mimic graphs. Sec.~\ref{sec:conclusions} concludes the paper with a discussion of the results and an outlook on future research.

\section{The model}\label{sec:models}
In this section we discuss the spatial search by CTQW and the geometric renormalization technique for complex networks. Numerical methods used in the calculations of the results are also considered.

\subsection{Spatial search by CTQW}
Given a network described by an undirected graph $G=(V,E)$ with $|V|=N$ nodes, a set $E$ of edges and no self-loops, the adjacency matrix $A$ is defined as
\[A_{jk} = \begin{cases}
1,\ (j,k) \in E \\
0,\ (j,k) \notin E,
\end{cases}\]
where $(j,k)$ describes an edge between nodes $j$ and $k$. We define the \emph{Laplacian} matrix of the graph as $L = D-A$, where $D$ is the diagonal matrix $D = \text{diag} (k_1, ..., k_N )$, $k_i$ being the degree of node $i$, $k_i = \sum_j A_{ij}$.

The spatial search algorithm is aimed at finding a given node $w$ of the network by considering a CTQW on the $N$-dimensional Hilbert space spanned by the basis states $\{\ket{1}, \ldots, \ket{N}\}$. The walker is prepared in some initial state $\ket{\psi(0)}$ and then letting it evolve according to the Hamiltonian
\begin{equation}\label{Hamiltonian}
    H = \gamma L + H_w,
\end{equation}
where $H_w = - |w\rangle \langle w|$ is the \emph{oracle Hamiltonian}, which causes the amplitude to accumulate to the target site $|w\rangle$, and $\gamma$ is a suitable coupling constant that is independently chosen. The value of $\gamma$ regulates the rate at which the probability amplitude flows between nodes per unit time. We have chosen to neglect a constant in the oracle Hamiltonian with units of inverse time, which also makes $\gamma$ a dimensionless parameter.

The state of the walker at time $t$ is the solution of the Schrödinger equation,
\begin{equation*}
    i \frac d{dt} |\psi (t) \rangle = H |\psi (t)\rangle.
\end{equation*}
In spatial search, the initial state of the walker is usually set to be an equal superposition between all the nodes, i.e.
\begin{equation}\label{Initial}
    | \psi (0)\rangle = |s\rangle =  \frac{1}{\sqrt{N}} \sum_j |j \rangle.
\end{equation}
The state of the walker at time $t$ is then
\begin{equation}\label{State}
    |\psi (t) \rangle = e^{-iHt} |s\rangle.
\end{equation}

The spatial search is successful if, when measuring the position of the quantum walker at a certain time $t$, we find the walker in the target state $\ket{w}$ with high probability, the probability being $p_w (t) = |\langle w | \psi (t) \rangle |^2$. One thus needs to optimize this probability with respect to both $t$ and $\gamma$ so that the success probability becomes as close to one as quickly as possible. These optimized values of $t$ and $\gamma$, namely $T$ and $\gamma_{\text{opt}}$, respectively, may in general depend on properties of the chosen target state and of the network. 

We write the success probability as
\begin{equation}\label{Success probability}
    p_{succ} = \max_t p_w(t) ,
\end{equation}
and indicate with $T$ the smallest value of $t$ that optimizes the success probability. We say that the search algorithm is optimal if $p_{succ} = \mathcal{O} (1)$ in time $T = \mathcal{O} (\sqrt{N})$, where $N$ is the number of nodes in the graph. When the success probability is not close to one, multiple repetitions of the algorithm may be necessary in order to have a high confidence of finding the target node.
One should thus look at the search time, defined as \cite{Cattaneo}
\begin{equation}\label{eq:search_time}
    T_w = \frac{T} {p_{succ}}.
\end{equation}

\begin{figure*}
    \centering
    \includegraphics[width=0.9
    \textwidth]{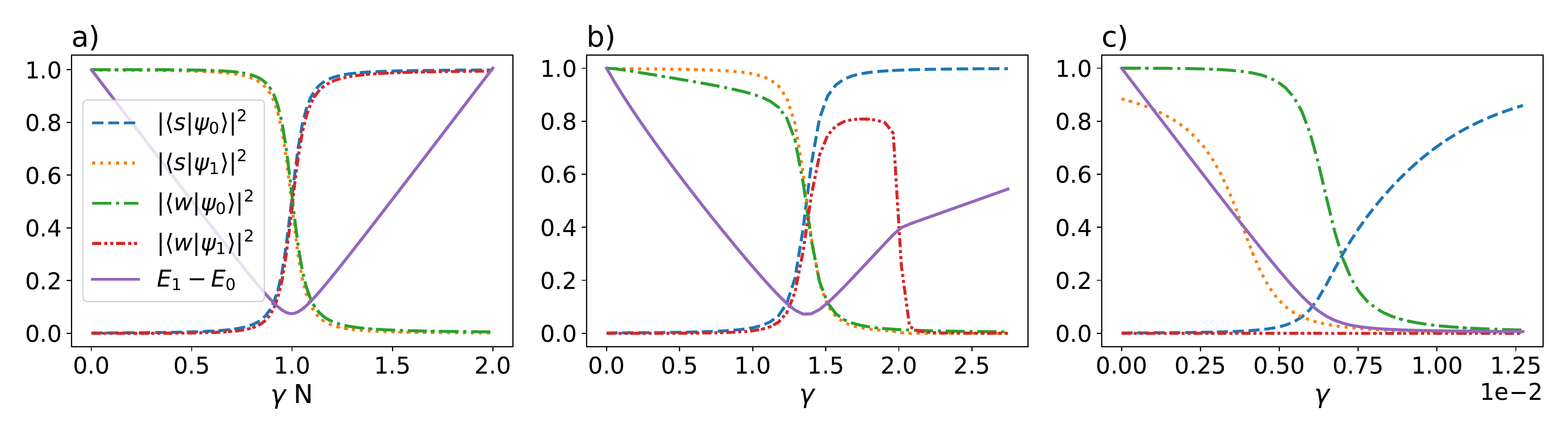}
    \caption{The overlaps of the ground and first excited states ($\ket{\psi_0}$ and $\ket{\psi_1}$) of the Hamiltonian with the initial and target states ($\ket{s}$ and $\ket{w}$, respectively), as well as the differences between the ground ($E_0$) and excited state energies ($E_1$) in different scenarios. Plot a) shows the case for the complete graph, where a clear point for the largest overlap can be seen for $\gamma = 1/N$. Plot b) corresponds to a node of degree 1 and plot c) to the highest-degree node in the smallest replica of the Internet network considered in this paper.}
    \label{fig:overlaps}
\end{figure*}

Following Ref.~\cite{ChildsGoldstone} we recall that, for any graph, $\vert s \rangle$ is the ground state  of the Laplacian, and therefore, for $\gamma \rightarrow \infty$ the ground state $\vert \Psi_0\rangle$ of the Hamiltonian in Eq.~\eqref{Hamiltonian} is close to $\vert s \rangle$. On the other hand, for $\gamma \rightarrow 0$, $\vert \Psi_0\rangle$ is close to the target state $\vert w \rangle$. Moreover, since the overlap between initial and target state is inversely proportional to $\sqrt{N}$, degenerate perturbation theory shows that the first excited state $\vert \Psi_1\rangle$ will be close to $\vert s \rangle$ for $\gamma \rightarrow 0$ in the large $N$ limit. For many graphs, there exists some intermediate range of values of $\gamma$ for which the ground state changes from target to initial state, and at the same time the states $\vert \Psi_0\rangle$ and $\vert \Psi_1\rangle$ have substantial overlap with both $\vert w \rangle$ and $\vert s \rangle$. Under these conditions, the Hamiltonian  drives transitions between these two states, and thus rotates the initial state $\vert s \rangle$ to a state with substantial overlap with the target state in a time inversely proportional to the energy gap $E_1-E_0$, where $E_0$ is the ground state energy and $E_1$ is the first excited state energy. As we can see from Fig.~\ref{fig:overlaps} a), for the complete graph, this is indeed what happens. In this case there exists a value of $\gamma$ for which the energy gap approximately closes and almost perfect oscillations between the initial and target states occur. In this case, the spatial search is optimal, showing a quadratic speedup \cite{ChildsGoldstone}. 

Real networks lack the symmetry of complete graphs and have very different topologies, so we expect dependence on the target node. To begin grasping the main properties of spatial search on real networks, we show in Fig.~\ref{fig:overlaps} b) and c) the overlaps of the lowest two eigenstates of the search Hamiltonian with both the target and the initial state, as well as the energy gap, for a renormalized replica of the Internet and two exemplary nodes, namely a node with degree one and the node with the largest degree. The behaviour of the plotted quantities for the node of degree one retains some features seen in the complete graph, while the case of the highest degree node shows clear differences. Specifically, it is clear that, for increasingly large values of $\gamma$ the first excited eigenstate does not coincide with the target state any longer, so while the gap closes, the dynamics are not anymore confined to the two-dimensional subspace generated by $\vert s \rangle$ and $\vert w \rangle$ and we expect the probability of transition to the target state to be significantly reduced.

In this paper we will find the optimal values $\gamma_{opt}$ and $T$ numerically. We will use the approximate values derived in Ref.~\cite{optimality} as one set of the initial conditions for the optimization, as well as a benchmark for our results. Following the derivation of Ref.~\cite{optimality} with our choice of Hamiltonian given in Eq.~(\ref{Hamiltonian}), we derive the initial approximate optimal values for $\gamma$ and time. 
We can recast the search Hamiltonian as $H_{\text{search}} = \gamma_{\text{norm}} H_{\text{norm}} - H_w$, where $H_{\text{norm}} = I - L / \lambda_L$ is a normalized Hamiltonian. Here $\lambda_L$ is the largest eigenvalue of $L$. The eigenvalues of $H_{\text{norm}}$ now lie in the interval $[0,1]$ as $0 = \lambda_1 \leq \lambda_2 \leq ... \leq \lambda_N = 1$ with the corresponding eigenstates of $\lambda_N$ and $\lambda_{N-1}$ being the ground and first excited states of $L$, respectively. 
The spectral gap is given by $\Delta = 1- \lambda_{N-1}$. The eigenvalues $\lambda_i$ and corresponding eigenstates $|v_i \rangle$ of $H_{\text{norm}}$, i.e. $H_{\text{norm}} |v_i\rangle = \lambda_i |v_i \rangle$, can be used to express first the target node in the basis of the eigenstates as
\begin{equation}\label{basisofeigenstates}
    | w \rangle = \sum_{i=1} ^N a_i |v_i \rangle
\end{equation}
and then to define the parameters
\begin{equation}\label{S_k parameters}
    S_k = \sum_{i=1} ^{N-1} \frac{|a_i |^2}{(1-\lambda_i )^k}.
\end{equation}
The approximate optimal solutions for $\gamma$ and $T$ derived for generic networks in Ref. \cite{optimality}
are given by
\begin{equation}\label{approximateoptima}
    \gamma \approx S_1 / \lambda_L, \qquad T = \mathcal{O}\left( \frac{1}{\sqrt{\epsilon}} \frac{\sqrt{S_2}}{S_1} \right),
\end{equation}
where $\epsilon = 1 / N$. 
The authors also impose a \emph{spectral condition} for the regime of validity in which the approximations are accurate:
\begin{equation}\label{spectralcondition}
    \sqrt{\epsilon} \leq c \min \Big\{ \frac{S_1 S_2}{S_3}, \Delta \sqrt{S_2} \Big\},
\end{equation}
where $c$ is a sufficiently small positive constant.

\subsection{Geometric renormalization of complex networks}
\label{sec:geometric_renormalization}
In this subsection we give a short overview of the state of the art and key concepts in the geometric renormalization of complex networks.

Despite their inherent complexity, real-world networks belonging to different domains, from the Internet to metabolic networks, exhibit certain common characteristic traits, such as degree heterogeneity (that is, most nodes have a low degree while a few nodes are connected to a macroscopic fraction of the system), high levels of clustering (high density of connected triples of nodes), and the small-world property (pairs of nodes lie a few hops away from each other), among others.

Naturally, since the early days of network theory, one of the main goals has been to explain these properties across domains.
Yet, the origin of clustering remained elusive until the proposal of so-called geometric models \cite{self-similarity}.
In this theoretical framework, clustering is considered to be a reflection of similarity between nodes, that is,
nodes of relatively low \textit{popularity} (low-degree nodes) can nevertheless be connected to one another with high probability if they share some other common traits.

For example, in the Internet, similarity space is found to be congruent with geographic distance \cite{sustaininginternet}.
In the paradigmatic network geometric models, such as the $\mathbb{S}^1$ \cite{self-similarity} and the $\mathbb{H}^2$ \cite{hyperbolicgeometry}, similarity space is mathematically represented in terms of an underlying metric space (often one-dimensional for simplicity), and the connection probability between nodes depends on the distance between them in the metric space.
Interestingly, these models enable the so-called embedding of the network: one can find the coordinates of the nodes of a real network in the abstract similarity space by means of a likelihood maximization approach \cite{sustaininginternet, mercator}.
The resulting network embeddings not only reveal the underlying similarity between nodes, but they can even be used to navigate the network efficiently \cite{sustaininginternet,geometricrenormalization}.

Network geometric models excel at generating networks that resemble real-world networks with remarkable accuracy, including the prediction of highly non-trivial topological properties of real systems (see \cite{networkgeometry} for a recent review).
One such result is the prediction that real-world graphs must be self-similar with respect to a coarse-graining transformation in similarity space, which was confirmed in real systems \cite{geometricrenormalization}.

This discovery was further exploited to develop a technique to produce \textit{smaller-scale replicas} of real networks, which have very similar topologies to the original networks despite their reduced size.
In Ref.~\cite{geometricrenormalization}, it was also found that dynamical processes running on the replicas exhibit the same behaviour as in the original network, which enables not only to study dynamics on smaller substrates (with reduced computational costs), but also to analyze the dependence of dynamical processes on the system size even for real-world systems, for which one typically only has access to a single, fixed-size instance \cite{guillepnas}.

While the details of the renormalization can be found in Ref.~\cite{geometricrenormalization}, we give here an intuitive explanation of how it works. Once the network has been embedded in a geometric space, we define groups of $r$ nodes with close coordinates in the metric space, and we then replace them with \emph{supernodes}. Each supernode is then placed within the angular region defined by the corresponding block so that the order of nodes is preserved. A link between two supernodes is added if there are one or more links between the underlying groups of nodes.  This operation can be iterated to obtain many renormalized network layers $l$, starting from the original network (which we identify as layer $l=0$). Each layer $l$ is then $r^l$ times smaller than the original one. Here we set $r=2$, that is, we group nodes in pairs.
This transformation produces smaller networks with similar topological properties except for the average degree. In fact, the renormalization flow of the average degree signals the small-worldness of the network under consideration. To obtain smaller-scale replicas on which dynamical processes resemble those running on the original network, links must therefore be removed to match the original average degree. This is done stochastically according to the underlying $\mathbb{S}^1$ model in a so-called \textit{pruning} procedure that preserves the topological features of the graph \cite{geometricrenormalization}.

In this paper, we exploit this technique to analyze the efficiency of spatial search by CTQW on the Internet autonomous system network \cite{Internetnetwork, internetmapping} as a function of the system size, hence addressing the question of whether spatial search is optimal for real-world communication network topologies.

\subsection{Numerical methods}
A geometric embedding algorithm for networks is used in order to obtain coordinates in a hidden metric space for the network nodes, and to then renormalize the network according to the method described in Sec. \ref{sec:geometric_renormalization}. The algorithm is available in the open-source package Mercator \cite{mercator}, which allows us to embed the complex networks in an $\mathbb{S}^1 / \mathbb{H}^2$ space. 
Mercator uses a combination of Laplacian eigenmaps (LE) and maximum likelihood (ML) estimation methods to
find the coordinates and other model parameters of every node in the underlying metric space for which the congruency between the $\mathbb{S}^1$ model and the observed real-world complex network is maximized. The $\mathbb{S}^1$ model can be mapped to the $\mathbb{H}^2$ model, which lives in a disk in hyperbolic space. In Fig.~\ref{fig:visualization}, we use the $\mathbb{H}^2$ representation to depict the networks.

The optimal values of the success probability, Eq.~\eqref{Success probability}, in terms of search time $t$ and the parameter $\gamma$ are obtained via numerical maximization, starting from two possible initial guesses: one is the approximate solution of Eq.~\eqref{approximateoptima}, while the other one is $t = 0$ and $\gamma = 1 / k$, where $k$ is the degree of the target node. A standard implementation of the gradient-free Nelder-Mead algorithm is used in the optimization. These numerical optimizations are performed for each node as the target node separately in the case of each graph. For each node, the initial guess that gives the optimal result is chosen.

In principle, the calculation of the state of the walker, Eq. (\ref{State}), requires diagonalizing the Hamiltonian, which is a costly (or even unfeasible) task for the largest networks, even when sparse diagonalization algorithms are used. There are, however, algorithms that allow to efficiently evaluate the action of the exponential of a sparse matrix on a vector. One of such algorithms is Expokit, introduced in \cite{expokit}. The algorithm makes use of Krylov subspace projection methods in order to iteratively evaluate the action of a large sparse matrix exponential on a vector, without having to calculate the exponential directly.

\section{Results}\label{sec:results}
In this paper, we consider different renormalized replicas of a snapshot of the Internet network at the level of autonomous systems taken in 2009 \cite{Internetnetwork, internetmapping, geometricrenormalization}. The full network contains $23\,748$ nodes and $58\,414$ edges.
We show results that have been calculated for five different pruned layers of the Internet, and the number of nodes and edges can be seen in Table \ref{tab:table1}. Generally, the pruned networks are used to assess dynamics for different system sizes, while the renormalized ones tell us about the structure of the network at different scales (see \cite{geometricrenormalization} for details). Since we are interested in CTQW, we focus on the pruned replicas. Note that only the giant component is considered if the network has some disconnected components. In particular, the pruned Internet networks have from two to five disconnected components with 2-3 nodes each, which are left out. 

For clarity, in the rest of the paper we will use the term 'renormalized' in place of 'pruned', since the process of renormalization is the exact same in both cases, with only the stochastic pruning of the edges as an addition to the latter.

Erdös-Renyi (ER) \cite{Erdos-Renyi} and Barab{\'a}si-Albert (BA) \cite{Barabasi-Albert} graphs are also analyzed, to compare to the different renormalized replicas of the Internet. The synthetic graphs are generated in such a way that they have approximately the same number of nodes and edges of the Internet replicas \footnote{In particular, for the BA graphs, the number of nodes is exactly the same, while the number of edges is slightly different, since the generation model only allows for multiples of the number of nodes. The number of edges was then decided by looking at the average degree $\langle k \rangle$ of the Internet network and the rounding to the closest integer to get the multiple of the number of nodes. The number $3$ was chosen for all BA graphs since the average degree of the renormalized versions is around $2.5$ for all of them. The ER graphs were generated simply by generating a random graph with the same amount of nodes and edges as the corresponding Internet network. When generating ER graphs similar to the Internet replicas, the low average degree gives rise to many disconnected components. We chose to generate graphs with a higher initial number of nodes to offset the removal of the smaller disconnected components, so in these cases the generated graphs are not fully random.}.

\begin{table}[b]
\caption{\label{tab:table1}%
The number of nodes ($N$) and edges ($m$) in different layers of the Internet network replicas. Only the giant component is considered for each layer.
}
\begin{ruledtabular}
\begin{tabular}{lrr}
\textrm{Layer} & $N$ & $m$ \\
\colrule
l=1 &  11626 &  28880 \\
l=2 &  5729 &  14660 \\
l=3 &  2846 &  7439 \\
l=4 &  1409 &  3679 \\
l=5 &  697 &  1817 \\
\end{tabular}
\end{ruledtabular}
\end{table}

\begin{figure*}
    \centering
    \includegraphics[width=0.99 \textwidth]{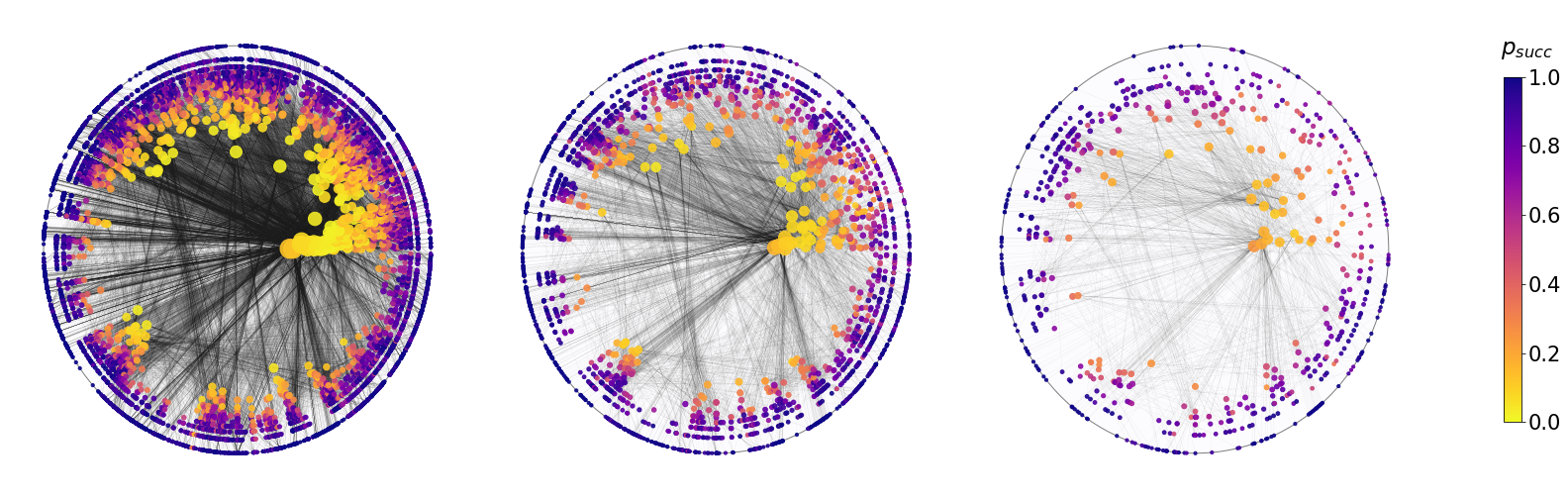}
    \caption{Visualizations of the Internet network replicas, layers $l=1$, $l=3$ and $l=5$ from left to right, in the hyperbolic embedding $\mathbb{H}^2$. In this embedding, the radial coordinate of a node is related to the logarithm of its degree, which is also represented by the size of the node. The color indicates the optimal success probability of the spatial search for a given target node. We notice
    that nodes with larger degrees (hubs) tend to have a lower success probability than small-degree nodes (see also Fig.~\ref{fig:Internet layers gammas sp degrees}).}
    \label{fig:visualization}
\end{figure*}

Figure~\ref{fig:visualization} shows three layers of renormalization of the Internet network in the $\mathbb{H}^2$ hyperbolic space, after pruning. The effect of the renormalization is easily visualized: the number of nodes and links is reduced, but the structure of the network is preserved. In the figure, the optimal probability of success for a given target node is indicated by the color, while the size indicates the degree; the larger the degree, the bigger the node. 
We can easily see that $p_{succ}$ is higher for small-degree nodes and much lower for large-degree nodes.

\begin{figure*}
    \centering
    \includegraphics[width=0.9 \textwidth]{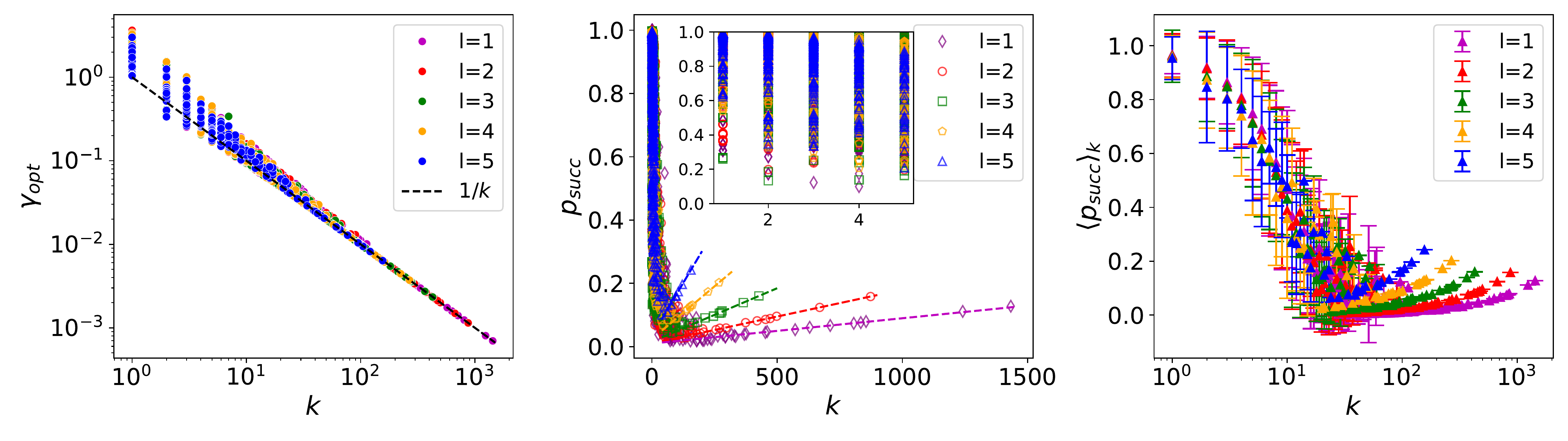}
    \caption{The left-most plot shows the optimal values of $\gamma_{opt}$ of the Internet replicas with respect to the degree $k$ in log-log scale. The dashed black line shows the function $1/k$, which is closely followed by $\gamma_{opt}$, especially for large-degree nodes, while there are some fluctuations for small-degree nodes. The middle plot shows optimal success probabilities $p_{succ}$ of the different layers of the renormalized Internet networks with respect to the degree $k$, as well as linear functions fitted to the  success probabilities of the larger-degree nodes The constants of the linear function are given in Table \ref{tab:spsdegreeslinearparts}. The inset shows a close-up of small-degree nodes. The right-most plot shows the averages of the optimal success probabilities over degree classes $\langle p_{succ}\rangle_k$, as well as their standard deviations.
    }
    \label{fig:Internet layers gammas sp degrees}
\end{figure*}

The left-most plot in Fig. \ref{fig:Internet layers gammas sp degrees} shows the optimal values $\gamma_{opt}$ of the search parameter for different layers of the renormalized Internet networks as a function of the degrees, in logarithmic scale. The values of $\gamma_{opt}$ follow the $1/k$ relation quite closely, with more variance in the low-degree nodes of larger networks. 
The middle panel of Fig. \ref{fig:Internet layers gammas sp degrees} shows the optimal success probability versus the degree of all nodes for the various layers of the Internet network. The inset highlights the success probabilities in the small-degree region.
This panel shows that for a small-degree node $p_{succ}$ tends to be larger on average, and for large-degree nodes it tends to be smaller. This can be intuitively understood by considering that if the target state is a small-degree node, then in the Hamiltonian $H= \gamma L - |w \rangle \langle w|$ the oracle causes a more significant change to the corresponding diagonal element (since we consider the dynamics in the node basis), than if the target state was a large-degree node. Correspondingly, the largest-degree nodes, or more accurately their states, correspond almost exactly to eigenstates of the Laplacian. 

The fact that spatial search algorithm is more optimal for the peripheral small-degree nodes than for the hubs is in contrast to the case of the star graph: there, the central node is the easiest to find for the algorithm \cite{Cattaneo}. However, the case of the star graph is anomalous in that it is highly symmetric, with one central node, and with all other nodes being connected only to that node. In this case with the spatial search, regardless of whether the target is the central or an outer node, the eigenspectrum of the search Hamiltonian is $N-2$ degenerate, which greatly affects the dynamics of the walk in comparison to complex networks. Furthermore, in the computational limit $N>>1$ there is no significant difference for the optimality of the search in the two different types of targets in the star graph \cite{Cattaneo}.

For the nodes with the largest degrees, there seems to be a linear dependence between $p_{succ}$ and the degree. To assess this dependence quantitatively, we fitted a linear function $p_{succ} (k) = a k + b$ to the data points of the largest degree nodes for each layer. The coefficients are reported in Table~\ref{tab:spsdegreeslinearparts}. We can see that the slope nearly doubles between consecutive layers.
Figure \ref{fig:Internet layers gammas sp degrees} shows that for nodes with intermediate connectivity there is no clear dependence between degree $k$ and success probability $p_{succ}$. We investigated the possible dependence of $p_{succ}$ on other quantities related to the local topology of nodes, such as the weighted degree, eigenvector centrality, and the local clustering coefficient, but a clear connection was not found in any of these cases.

The right-most plot in Fig.~\ref{fig:Internet layers gammas sp degrees} shows the averages of optimal success probabilities with the same degree and their standard deviations in logarithmic scale. We also checked that the scaling of the optimal time $T$ is $\mathcal{O}(\sqrt{N})$ for all nodes with $p_{succ}$ of the order $\mathcal{O}(1)$ (results are not shown here). Later in the paper we show that the scaling of the optimal time when taking the average is also $\mathcal{O}(\sqrt{N})$. This highlights the fact that for small-degree nodes the search is always optimal, and only sub-optimal for mid- to large-degree nodes, which make up only a small fraction of all the nodes in the network.

\begin{table}[b]
\caption{\label{tab:spsdegreeslinearparts}%
The constants for the fitted linear functions of the optimal success probabilities of the Internet network replicas given with respect to the degree as seen in the middle plot of Fig. \ref{fig:Internet layers gammas sp degrees}.}
\begin{ruledtabular}
\begin{tabular}{l|cc}
 & \multicolumn{2}{c}{\textrm{Renormalized}} \\
 \colrule
 $p_{succ} = ak + b$ & a & b \\
 \colrule
 $l=1$ & $0.8 \times 10^{-4}$ & 0.010 \\
 $l=2$ & $1.6 \times 10^{-4}$ & 0.014 \\
 $l=3$ & $2.8 \times 10^{-4}$ & 0.018 \\
 $l=4$ & $6.2 \times 10^{-4}$ & 0.025 \\
 $l=5$ & $10.0 \times 10^{-4}$ & 0.062
\end{tabular}
\end{ruledtabular}
\end{table}

\begin{figure}
    \centering
    \includegraphics[width=0.9\columnwidth]{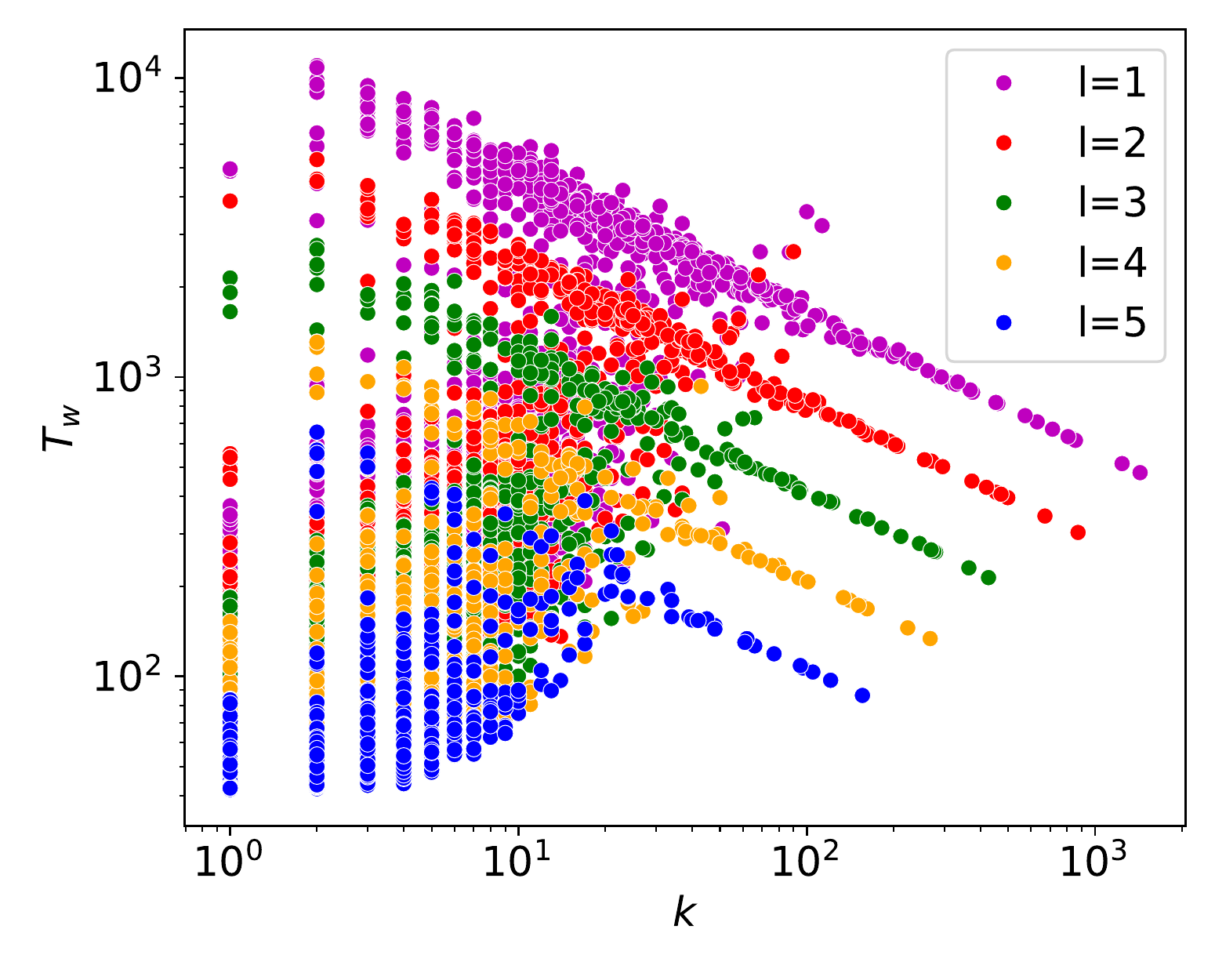}
    \caption{Log-log plot of the optimal search time $T_w$ versus the degree $k$ for the nodes of the five renormalized layers of the Internet network. As one can expect, the optimal search time increases with the size of the network. For the largest-degree nodes, the search time $T_w$ and the degree $k$ are correlated, with $T_w \sim k^{-1/2}$.}
    \label{fig:searchtimes}
\end{figure}

Let us now look at the optimal search times, and how they depend on the characteristics of the target node. Figure \ref{fig:searchtimes} shows the optimal search times for all nodes of different layers of renormalization of the Internet, given as a function of the degrees in logarithmic scale. As one would expect, in general, the smaller the size of the replica, i.e., the larger the renormalization layer index $l$, the shorter the optimal search time. For the largest-degree nodes, the search time $T_w$ and the degree $k$ are correlated, as it was for the success probabilities depicted in Fig.~\ref{fig:Internet layers gammas sp degrees}. However, the scaling is now given by $T_w \sim k^{-1/2}$. One would then expect, for consistency, the optimal time $T$ of the largest-degree nodes to scale roughly as $T \sim k^{1/2}$, which is indeed the case (not shown).

\begin{figure}
    \centering
    \includegraphics[width=0.9
    \columnwidth]{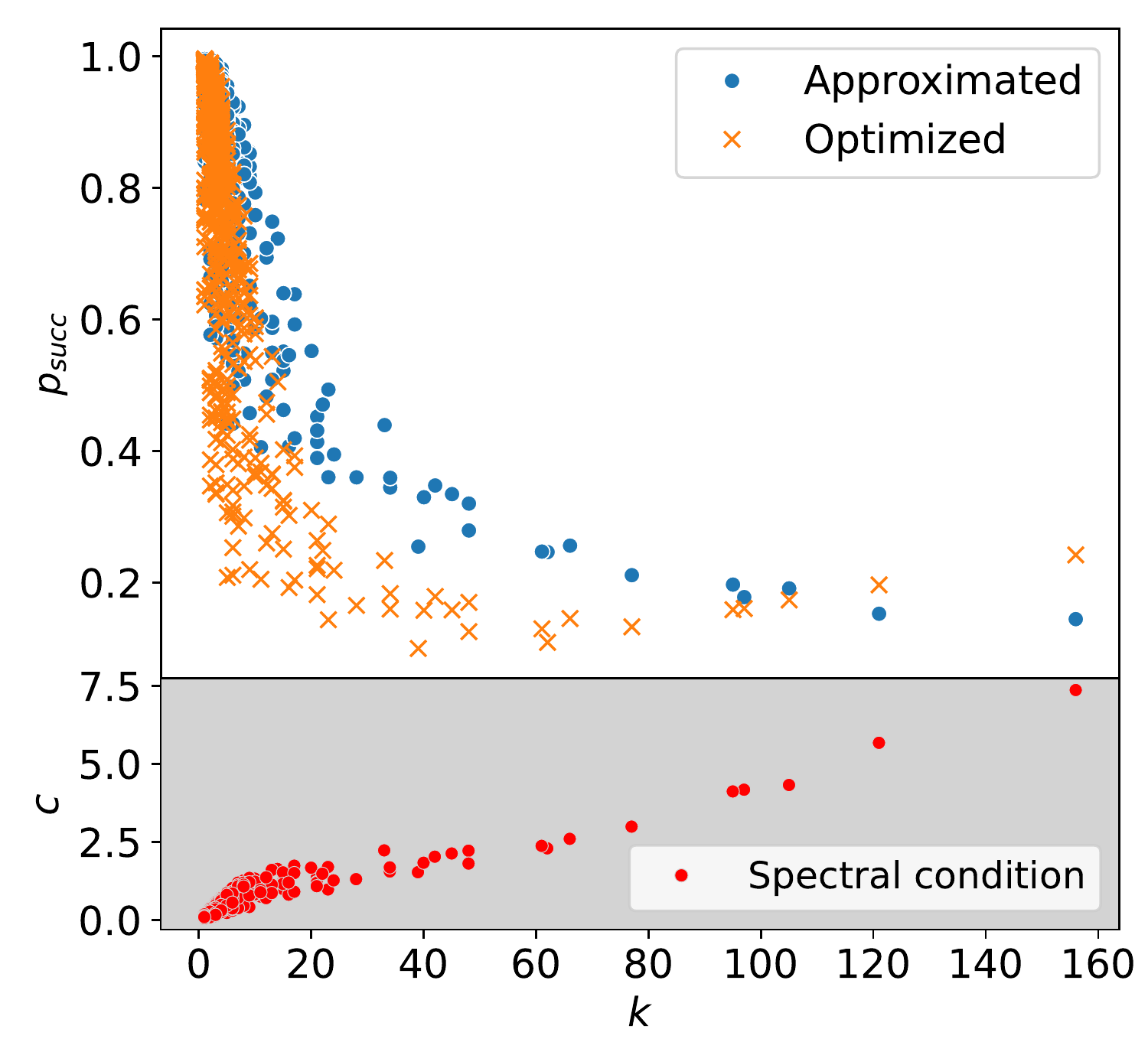}
    \caption{The calculated approximated and optimized values of optimal success probabilities for the $l=5$ network, as well as the constant $c$ with respect to the degree, which indicates a decrease in the accuracy of the prediction.}
    \label{fig:Internet optimized vs approximated}
\end{figure}

We now assess to which extent the approximated results of Eq.~\eqref{approximateoptima} capture the main features of the quantum spatial search on a real network like the Internet, or more specifically its smaller-scale replicas. Figure \ref{fig:Internet optimized vs approximated} compares the approximated values of $p_{succ}$ obtained with Eq.~\eqref{approximateoptima} to the numerical results for layer $l=5$ of the renormalized Internet network. From the plot it is evident that the approximation is quite accurate for small-degree nodes, but deviates from the numerical optimum for large-degree nodes. This can be explained by calculating the value of the constant $c$ from Eq. (\ref{spectralcondition}), which is shown in the lower part of the plot; we notice that $c$ is almost linearly dependent on $k$. As a consequence, the accuracy of the approximation decreases as $k$ increases. Indeed, in the upper plot we do see that the approximate and optimized values overlap well only for small-degree nodes. Although the two values are very similar in the region of nodes with degrees roughly $100$, the slopes are different, so the approximate solutions do not predict the overall behavior for large-degree nodes very well.

\begin{figure*}
    \centering
    \includegraphics[width=1.0\textwidth]{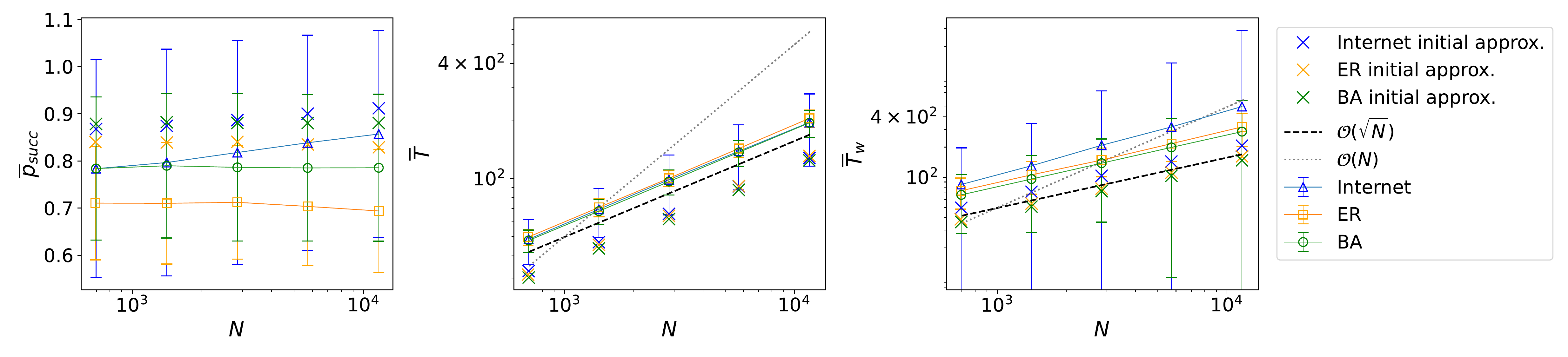}
    \caption{The averages and standard deviations of optimal success probabilities, times and search times of the renormalized Internet networks, as well as those of the ER and BA graphs, as functions of the number of nodes $N$. The optimal quantum scaling $\mathcal{O} (\sqrt{N})$ is highlighted by a black dashed line, and the classical scaling $\mathcal{O} (N)$ is highlighted by a gray dotted line. The lines show the same quantities calculated using the approximations in Eq.~\eqref{approximateoptima}. }
    \label{fig:avgstimessps}
\end{figure*}

\begin{table}[b]
\caption{\label{tab:fittedvalues}%
The constants for the fitted logarithmic function ($\log \overline{T}_w = c + x \log N $) on the average search times of the Internet network replicas as well as the ER and BA mimic graphs as seen on the right-most plot of Fig.~\ref{fig:avgstimessps}. For the exponential constant $x$ the standard error is also provided.}
\begin{ruledtabular}
\begin{tabular}{l|cc|cc}
 & \multicolumn{2}{c}{\textrm{All nodes}} & \multicolumn{2}{c}{\begin{tabular}{@{}c@{}}$99~\%$ of nodes in order \\ of ascending degree\end{tabular}} \\
 \colrule
 \begin{tabular}{@{}c@{}}$\log \overline{T}_w =$ \\ $c + x \log N $\end{tabular} & c & x & c & x \\
 \colrule
 Internet & 0.290 & $0.633 \pm 0.006$ & 1.115 & $0.511 \pm 0.015$ \\
 ER & 0.914 & $0.516 \pm 0.007$ & 0.915 & $0.516 \pm 0.006$\\
 BA & 0.838 & $0.514 \pm 0.001$ & 0.864 & $0.504 \pm 0.009$
\end{tabular}
\end{ruledtabular}
\end{table}

As noted in Sec.~\ref{sec:introduction}, when assessing the performance of spatial search, one is interested in the dependence of the search time and success probability as functions of the size of the graph. In the ideal case, the search time will grow as $\mathcal{O}(\sqrt{N})$. Since in non-regular graphs the search time and probability depend on the individual target node, we look at these quantities averaged over all the nodes of the network. The results are shown in Fig.~\ref{fig:avgstimessps} for the various layers of the Internet replicas, as well as for ER and BA graphs of similar sizes. 

We  see that the average success probability $\overline{p}_{succ}$ for the renormalized Internet graphs increases with increasing $N$, which indicates that when the average degree is held constant while scaling down the network, the optimal success probability of the walker gets worse on average. 
In the case of the ER and BA graphs, on the other hand, $\overline{p}_{succ}$ remains constant as a function of $N$.
Hence, the average success probability of spatial search on a real network scales better with the system size than on equivalent synthetic graphs.

The average optimal time $\overline{T}$ for all graphs follows the optimal $\mathcal{O}( \sqrt{N})$. 
The averages of the search times $\overline{T}_w$, defined in Eq.~\eqref{eq:search_time}, increases more rapidly with increasing $N$ for the Internet networks.
The synthetic graphs mimicking the network replicas on the other hand, show almost optimal scaling, being very close to $\mathcal{O}(\sqrt{N})$. To substantiate these interpretations of the plots, we fitted a logarithmic function to the data points, which we discuss in detail in the last two paragraphs of this section. When this figure of merit is taken into account, the performance of spatial search on a real-world complex network is better than classical, but not optimal. Note that the average times $\overline{T}$ have a similar scaling for the Internet and mimic replicas while the success probability is on average higher for the renormalized Internet. Nonetheless, $\overline{T}_w$ for the Internet replicas is  always larger than for the mimic ER and BA graphs. Recalling Eq.~\eqref{eq:search_time}, this suggests a non-trivial correlation between success probability and optimal search time for the nodes in the network. Indeed, we find an anti-correlation between these two quantities that increases with the system size. Under these circumstances, it is reasonable to expect $\overline{T}_w$ to become increasingly larger than $\overline{T} / \overline{p}_{succ}$ as $N$ increases.

The averages of the approximated values in Fig.~\ref{fig:avgstimessps} show that for all graphs the approximations overestimate the efficiency of the spatial search algorithm, since the approximated $\overline{p}_{succ}$ is always larger than the actual one, and the approximated $\overline{T}$ and $\overline{T}_w$ are always lower. The difference between the averages of approximated values and the actual values remains the same with changing $N$, for all the plotted quantities, with the exceptions of $\overline{T}_w$ and $\overline{p}_{succ}$ of the Internet replicas, for the first of which the averages of approximated values scale much better than the corresponding average calculated values, and for the latter where for smaller $N$ the difference is larger than for larger $N$. The better scaling for approximated $\overline{T}_w$ is caused by the approximate values overestimating the optimal success probabilities for the mid-degree nodes, which also have the largest search times. 

The standard deviations in Fig.~\ref{fig:avgstimessps} also show that the optimal times are quite uniform, the optimal success probabilities have more variance but the standard deviations stay similar between layers, and the standard deviations in the search times are much greater due to the standard deviations in both $\overline{p}_{succ}$ and $\overline{T}$. We can also see that the variance in Internet replicas is larger in all cases than the variance in ER and BA mimic graphs. For $\overline{p}_{succ}$ the changes in averages between layers are also always within the standard deviations, so not much can be interpreted about the scaling of $\overline{p}_{succ}$ from this.

To determine the scaling of $\overline{T}_w$, we fitted a logarithmic function of the form $c + x \log N$ to the logarithm of the data points of the optimized results. The values for the constants can be seen in Table~\ref{tab:fittedvalues}. The obtained exponent is $x \approx 0.63$ for the Internet replicas, so the scaling is indeed better than classical, but not optimal. For the ER and BA graphs the scaling exponents are approximately $0.52$ and $0.51$, respectively, so the spatial search is already almost optimal for those graph types.

We recall that the approximated quantities well describe the case in which the target state is a node with a small degree, lying at the periphery of the hyperbolic graphs of Fig.~\ref{fig:visualization}.
One may then argue that, if one is interested in searching such nodes---or equivalently in efficient transport by CTQW to such nodes---one observes a quadratic speedup, with optimal scaling $\mathcal{O}(\sqrt{N})$. Indeed, if we only take into consideration the first $99\%$ of the nodes in each network arranged in the order of ascending degree (essentially only discarding the hubs), and fit the logarithmic function again to the average optimized search times, then the scaling for the Internet replicas becomes approximately $N^{0.51}$, a noticeable improvement and very close to optimal (indeed, within one standard deviation). For the ER and BA graphs the scaling remains practically the same as in the previous case.

\section{Conclusions}\label{sec:conclusions}
In this article, we studied the efficacy of spatial search by CTQW on a real-world complex network, namely the Internet. Given the large size of the graph, its non-trivial structure and spectral properties, exact analytical solutions are not possible and numerical simulation is computationally intensive for full-sized graphs. We thus applied for the first time in the quantum information field a recently introduced geometric renormalization technique for complex networks to bring the graph to a tractable size, while also preserving its structural properties.
This also allowed us to study the scaling of spatial search with the size of the network.

We found numerically the success probability, optimal coupling constant and optimal search time for each node of various renormalized versions of the Internet, possessing
between $700$ and $12\,000$ nodes, approximately. We showed that the optimal value of the coupling constant is inversely proportional to the degree of the target node. The success probability is, in general, larger and close to one for small-degree target nodes, it decreases abruptly for nodes of intermediate degree, and grows linearly with the degree for large-degree nodes. 
We also found that recent approximate results \cite{optimality} on the success probability and search time do not hold well for the Internet and its renormalized replicas, in particular for large-degree nodes, and in general tend to overestimate the performance of the algorithm.

Finally, we found that the scaling of the average optimal time with the number of nodes in the network is quite close to the ideal $\mathcal{O}(\sqrt{N})$, showing advantage over classical search algorithms. However, if one does take into account that the average success probability is significantly smaller than one, and thus factors in the time required to repeat the search for a sufficient number of times, the scaling reduces to approximately $\mathcal{O}(N^{0.63})$. In general, this is due to the fact that the performance of the spatial search depends sensibly on the choice of the target node, and some of the nodes, such as the hubs in the Internet graph, show very low success probability. Indeed, the scaling gets much closer to the optimal one if we disregard $1\%$ of the nodes with the largest degrees.

In this work, we have considered the Childs and Goldstone algorithm for spatial search, where the state preparation does not assume knowledge of the topology of the graph. Other refined algorithms, such as the one recently introduced in \cite{apers2021quadratic}, could give better performance, although they generally require preparing the walker in an optimized state, which usually requires prior knowledge of the topology. 

Complex network science combined with quantum physics is a very new field with much left to discover. In particular, in further research we could look at the quantum speedup of the search time with respect to the hitting time of a classical random walk, rather than assuming a uniform search time $\mathcal{O}(N)$, since different nodes in these complex networks can have very different hitting times. In this paper we opted to follow the approach that is customary in the literature for checking the quantum speedup of the Childs and Goldstone quantum spatial search algorithm \cite{ChildsGoldstone}. 

Another interesting aspect to research is the performance of the algorithm with multiple marked nodes \cite{multiplemarkedvertices} on real-world network. Since no analytical bounds are currently known for the Childs and Goldstone algorithm when there are multiple marked nodes, this could offer valuable insight into the behavior of the spatial search algorithm.

Finally, a crucial topic to address in the future is the performance of spatial search by CTQW on complex networks in the presence of noise \cite{Cattaneo}. While in general detrimental for quantum information applications, noise has been shown to enhance transport for example in complex biological networks  \cite{Chisholm2021,Kurt2020}. 

The above aspects are particularly relevant in the context of quantum biology, and specifically interesting for the lively debate on whether quantum effects, described in terms of transport via CTQW and spatial search, may have a role in biological systems, including living organisms \cite{QuantumAspectsofLife}. In order to properly address this question one needs to analyze the dynamics on real biological networks. The tools and methods developed in this paper are ideally suited to such an endeavor.

In conclusion, we have shown how to meaningfully describe quantum spatial search, and more in general assess the performance of CTQW, on real-world complex networks.  In this sense our results bridge a gap between complex network theory and quantum science, and are therefore expected to lead to new pathways in quantum biology and communication, as well as quantum simulation and algorithms.

\acknowledgements{
The authors acknowledge financial support from the Academy of Finland via the Centre of Excellence program (Project No. 336810 and Project No. 336814). G.G.-P. acknowledges support from the Academy of Finland via the Postdoctoral Researcher program (Project no. 341985) S.M. and G.G.-P. acknowledge support from the emmy.network foundation under the aegis of the Fondation de Luxembourg. The computer resources of the Finnish IT Center for Science (CSC) and the FGCI project (Finland) are acknowledged. 
}

\bibliography{bibliography}

\end{document}